\documentclass[useAMS,usenatbib]{mn2e}

\usepackage{amssymb,amsmath,upgreek}

\usepackage{epsfig,graphicx,times,subfigure}

\newcommand{\msun}{M$_{\odot}$}
\newcommand{\rsun}{R$_{\odot}$}

\newcommand{\aaps}{A\&A}
\newcommand{\aj}{AJ}
\newcommand{\pasp}{PASP}
\newcommand{\apj}{ApJ}
\newcommand{\mnras}{MNRAS}

\newcommand{\apjs}{ApJS}
\usepackage{gensymb}
\def\deg{\hbox{$^\circ$}}

\title[AM CVn: Gaia14aae]
{Total eclipse of the heart: The AM CVn Gaia14aae / ASSASN-14cn}

\author[H. Campbell et al.]
{H. C. Campbell$^{1}$\thanks{E-mail:hcc@ast.cam.ac.uk},
	T. R. Marsh$^{2}$,
	M. Fraser$^{1}$,
	S.T. Hodgkin$^{1}$,	
        E. de Miguel$^{3,4}$,\newauthor
	B. T. G\"{a}nsicke$^{2}$,
	D. Steeghs$^{2}$, 
     	A. Hourihane$^{1}$,
	E. Breedt$^{2}$,
	S. P. Littlefair$^{5}$,\newauthor 
	S. E. Koposov$^{1}$, 
	 {\L}. Wyrzykowski$^{6,1}$,
	G. Altavilla$^{7}$, 
	N. Blagorodnova$^{1}$, 
	G. Clementini$^{7}$,\newauthor 
	G. Damljanovic$^{8}$,  		
	A. Delgado$^{1}$,
	M. Dennefeld$^{9}$, 
	A. J. Drake$^{10}$, 
	J. Fern\'andez-Hern\'andez$^{1}$,\newauthor 
	G. Gilmore$^{1}$,   
	R. Gualandi$^{7}$,
	A. Hamanowicz$^{6}$, 
	B. Handzlik$^{6}$,
	L. K. Hardy$^{5}$,\newauthor 
	D. L. Harrison$^{1,11}$,
      	K. I{\l}kiewicz$^{6}$, 
      	P. G. Jonker$^{12,13}$,    
      	C. S. Kochanek$^{14,15}$,\newauthor   
      	Z. Ko{\l}aczkowski$^{16}$, 
	Z. Kostrzewa-Rutkowska$^{6}$, 		   
       R. Kotak$^{17}$,
	G. van Leeuwen$^{1}$,
	G. Leto$^{18}$,  \newauthor
	P. Ochner$^{19}$,  
	M. Pawlak$^{6}$, 
	L. Palaversa$^{20}$, 
	G. Rixon$^{1}$,	 
	K. Rybicki$^{6}$,
	B. J. Shappee$^{21}$,\newauthor
	S. J. Smartt$^{17}$, 
	M. A. P. Torres$^{12,13}$,
	L. Tomasella$^{19}$,  
	M. Turatto$^{19}$,
	K. Ulaczyk$^{6,2}$,\newauthor
	S. van Velzen$^{22}$,  
	O. Vince$^{8}$,
	N. A. Walton$^{1}$,
	P. Wielg\'orski$^{6}$,
	T. Wevers$^{13}$,
	P. Whitelock$^{23,24}$,\newauthor
	A. Yoldas$^{1}$,	
	F. De Angeli$^{1}$,
	P. Burgess$^{1}$,
	G. Busso$^{1}$, 
	R. Busuttil$^{25}$,
	T. Butterley$^{26}$,\newauthor
	K. C. Chambers$^{27}$,
	C. Copperwheat$^{28}$,
	A. B. Danilet$^{29}$, 
	V. S. Dhillon$^{5}$, 
	D. W. Evans$^{1}$, \newauthor
	L. Eyer$^{20}$,  
	D. Froebrich$^{30}$, 
	A. Gomboc$^{31}$, 
	G. Holland$^{1}$, 
	T. W.-S. Holoien$^{15}$, 	  
	J. F. Jarvis$^{25}$, \newauthor
	N. Kaiser$^{27}$,
	D. A. Kann$^{32}$,	 
	D. Koester$^{33}$,
	U. Kolb$^{25}$, 
	S. Komossa$^{34}$, 	
	E. A. Magnier$^{27}$, \newauthor
	A. Mahabal$^{10}$,
	J. Polshaw$^{17}$,
	J. L. Prieto$^{35,36}$,   
	T. Prusti$^{37}$,		 
	M. Riello$^{1}$,	
	A. Scholz$^{38}$, \newauthor
	G. Simonian$^{15}$, 
	K. Z. Stanek$^{15}$, 
	L. Szabados$^{39}$, 	 
	C. Waters$^{27}$, 
	R. W. Wilson$^{26}$\\
	$^{1}$Institute of Astronomy, University of Cambridge, Madingley Road, Cambridge, CB3 0HA, UK\\
	$^{2}$Department of Physics, University of Warwick, Gibbet Hill Road, Coventry, CV4 7AL, UK \\
	$^{3}$CBA (Huelva), Observatorio del CIECEM, Matalasca\~{n}as, E-21076 Almonte, Huelva, Spain\\
	$^{4}$Departamento de F\'{i}sica Aplicada, Universidad de Huelva, E-21071 Huelva, Spain \\
	$^{5}$Department of Physics and Astronomy, University of Sheffield, Sheffield S3 7RH, UK\\
	$^{6}$Warsaw University Astronomical Observatory, Al. Ujazdowskie 4, 00-478 Warszawa, Poland\\
	$^{7}$INAF, Osservatorio Astronomico di Bologna, I-40127 Bologna, Italy\\  
	$^{8}$Astronomical Observatory, Volgina 7, 11060 Belgrade 38, Serbia\\
	$^{9}$Sorbonne Universit\`es, UPMC Universite Paris 6 et CNRS, UMR 7095,  Institut d'Astrophysique de Paris, 98 bis Bd. Arago, F-75014 Paris, France.\\
	$^{10}$California Institute of Technology, 1200 E. California Blvd, CA 91225, USA\\
	$^{11}$Kavli Institute for Cosmology Cambridge, Madingley Road, Cambridge, CB3 0HA, U.K\\
	$^{12}$SRON, Netherlands Institute for Space Research, Sorbonnelaan 2, NL-3584 CA, Utrecht, The Netherlands\\
	$^{13}$Dept. of Astrophysics, IMAPP, Radboud University Nijmegen, Heyendaalseweg 135, 6525 AJ, Nijmegen, The Netherlands\\
  	$^{14}$Center for Cosmology and AstroParticle Physics (CCAPP), The Ohio State University, 191 W. Woodruff Ave., Columbus, OH 43210, USA\\
	$^{15}$Department of Astronomy, The Ohio State University, 140 West 18th Avenue, Columbus, OH 43210, USA\\
	$^{16}$Instytut Astronomiczny, Uniwersytet Wroc{\l}awski, Kopernika 11, 51-622 Wroc{\l}aw, Poland\\
	$^{17}$Astrophysics Research Centre, School of Mathematics and Physics, Queen's University Belfast, Belfast BT7 1NN, UK \\
	$^{18}$INAF - Osservatorio Astrofisico di Catania, Via Santa Sofia 78, I-95123 Catania, Italy\\
	$^{19}$INAF, Osservatorio Astronomico di Padova, I-35122 Padova, Italy\\
	$^{20}$Observatoire astronomique de l'Universit\'{e} de Gen\`{e}ve, 51 chemin des Maillettes, CH-1290 Sauverny, Switzerland\\	 
	$^{21}$Carnegie Observatories, 813 Santa Barbara Street, Pasadena, California 91101, USA\\
	$^{22}$Department of Physics and Astronomy, The Johns Hopkins University, Baltimore, MD 21218, USA\\	 
	$^{23}$South African Astronomical Observatory, P.O. Box 9, 7935 Observatory, South Africa\\
  	$^{24}$Astronomy, Cosmology and Gravity Centre, Astronomy Department, University of Cape Town, 7701 Rondebosch, South Africa\\
	$^{25}$Department of Physical Sciences, The Open University, Walton Hall, Milton Keynes MK7 6AA, UK\\
	$^{26}$Centre for Advanced Instrumentation, Department of Physics, University of Durham, South Road, Durham, DH1 3LE, UK\\	
	$^{27}$Institute for Astronomy, University of Hawaii at Manoa, Honolulu, HI 96822, USA\\
	$^{28}$Astrophysics Research Institute, Liverpool John Moores University, IC2, Liverpool Science Park, 146 Brownlow Hill, Liverpool, L3 5RF, UK\\
	$^{29}$Department of Physics and Astronomy, Washington State University, 1245 Webster Hall, Pullman WA 99164, USA\\
	$^{30}$Centre for Astrophysics \& Planetary Science, The University of Kent, Canterbury, Kent CT2 7NH, UK \\
	$^{31}$Faculty of Mathematics and Physics, University of Ljubljana, Jadranska ulica 19, 1000 Ljubljana, Slovenia\\
	$^{32}$Th\"uringer Landessternwarte Tautenburg, Sternwarte 5, D-07778 Tautenburg, Germany\\	
	$^{33}$Institut f\"ur Theoretische Physik und Astrophysik, University of Kiel, D-24098 Kiel, Germany\\
	$^{34}$Max-Planck-Institut f\"ur Radioastronomie, Auf dem H\"ugel 69, D-53121 Bonn, Germany\\
	$^{35}$N{\'u}cleo de Astronom{\'i}a de la Facultad de Ingenier{\'i}a, Universidad Diego Portales, Av. Ej\'ercito 441, Santiago, Chile\\   	
	$^{36}$Millennium Institute of Astrophysics, Santiago, Chile\\
	$^{37}$ESA, ESTEC, Keplerlaan 1, PO Box 299, 2200 AG Noordwijk, The Netherlands\\
	$^{38}$School of Physics and Astronomy, University of St Andrews, North Haugh, St Andrews KY16 9SS, UK\\	
	$^{39}$Konkoly Observatory of the Hungarian Academy of Sciences, Budapest, Hungary, PO Box 67\\
	    }

\begin{document}

\date{Submitted to Monthly Notices of the Royal Astronomical Society}

\pagerange{\pageref{firstpage}--\pageref{lastpage}} \pubyear{}

\maketitle
\newpage
\label{firstpage}
\newpage
\pagebreak
\begin{abstract}
We report the discovery and characterisation of a deeply eclipsing AM CVn-system, Gaia14aae (= ASSASN-14cn). Gaia14aae was identified independently by the All-Sky Automated Survey for Supernovae \citep[ASAS-SN;][]{Sha:14} and by the {\it Gaia} Science Alerts project, during two separate outbursts. A third outburst is seen in archival Pan-STARRS-1 \citep[PS1;][]{Sch:12,Ton:12,Mag:13} and ASAS-SN data. Spectroscopy reveals a hot, hydrogen-deficient spectrum with clear double-peaked emission lines, consistent with an accreting double degenerate classification. We use follow-up photometry to constrain the orbital parameters of the system. We find an orbital period of 49.71 min, which places Gaia14aae at the long period extremum of the outbursting AM CVn period distribution. Gaia14aae is dominated by the light from its accreting white dwarf. Assuming an orbital inclination of 90${\deg}$ for the binary system, the contact phases of the white dwarf lead to lower limits of 0.78 \msun\ and 0.015 \msun\ on the masses of the accretor and donor respectively and a lower limit on the mass ratio of 0.019. Gaia14aae is only the third eclipsing AM CVn star known, and the first in which the WD is totally eclipsed. Using a helium WD model, we estimate the accretor's effective temperature to be $12900\pm200$~K. The three outburst events occurred within 4 months of each other, while no other outburst activity is seen in the previous 8 years of  Catalina Real-time Transient Survey \citep[CRTS;][]{Dra:09}, Pan-STARRS-1 and  ASAS-SN data. This suggests that these events might be rebrightenings of the first outburst rather than individual events.
\end{abstract}

\begin{keywords}
Binaries: eclipsing, Stars: Cataclysmic Variables 
\end{keywords}

\section{Introduction}

AM Canum Venaticorum (AM CVn) stars are a rare class of compact hydrogen-deficient interacting binaries, comprised of white dwarfs (WDs) accreting He-rich material from low mass degenerate or semi-degenerate companions \citep[see][for recent reviews]{Nel:05,Sol:10}. The orbital periods of these systems range from 5 to 65 min. This implies highly evolved components and makes them, along with their ultra-compact X-ray binary equivalents, one of the most compact classes of binary system known. The prototype system for the class of object was discovered in 1967, and has an orbital period of 17~min \citep{Sma:67,Pac:67}. Since then, 43 confirmed AM CVn systems have been discovered \citep{Lev:15}. Not only are these systems interesting as one of the possible end points for binary WD evolution \citep{Nel:01}, they are also potentially strong sources of gravitational wave emission due to their compact configurations \citep{Nel:03}, and they may be the progenitors of peculiar ``dot Ia'' supernovae \citep{Sol:05,Bil:07,Ins:15}.

As binaries, AM CVn systems can yield detailed information on the masses and radii of the two components if eclipses and radial velocity variations can be observed. Eclipsing systems in particular offer the possibility of measuring full system parameters, including inclination and component masses, from time-series photometry alone. The most robust results come from systems in which the white dwarf is totally eclipsed. The extreme mass ratios of AM CVns mean that the likelihood of observing such systems is low and currently only two eclipsing AM CVn systems are known. SDSSJ0926+3624 was the first eclipsing AM CVn star to be discovered \citep{And:05,Cop:11,Szy:14}, however its WD is only partially eclipsed. A second partial eclipser \citep[PTF1 J191905.19+481506.2,][]{Lev:14} was recently discovered, but it only eclipses the edge of the disc and not the WD, and so cannot be used for parameter determination.

Determining the nature of the secondary (donor) star is critical to our understanding of the past evolution of the system, since the three binary evolution channels proposed to form AM CVn stars are best distinguished by the state of the donor star at the onset of mass transfer. If the primary WD is accreting from another He-rich WD, the binary must have undergone two common envelope events in the past to reduce it to the observed compact configuration. This is known as the double degenerate channel \citep{Pac:67, Fau:72}. Alternatively, if the donor is not fully degenerate at the time when it leaves the second common envelope, the donor will be more massive than in the case of the double-degenerate channel. Mass loss will cause it to become increasingly degenerate as the binary evolves \citep{Sav:86,Ibe:87}. At the longest observed orbital periods (i.e., the oldest AM CVn systems), the two channels become indistinguishable. The donor is predicted to reach the same near-zero temperature, low entropy configuration in both cases \citep{Del:07}. A third possibility is that the binary may start mass transfer as a hydrogen-rich cataclysmic variable. Such a system could evolve to an AM CVn star if the donor star had already started to evolve by the time mass transfer starts, and results in a hotter, more massive donor star and traces of hydrogen may be expected in such systems \citep{Pod:03}.

In this paper, we present follow-up observations and preliminary modelling of the AM CVn system Gaia14aae \citep[RA = 16:11:33.97, Dec= +63:08:31.8,][]{Rix:14}. Gaia14aae was first detected in outburst by the All-Sky Automated Survey for Supernovae \citep[ASAS-SN,][]{Sha:14} at {\it V}~=~13.6 on 2014 June 14, who gave it the designation ASASSN-14cn. This was before the formal start of the {\it Gaia} Science Alerts project. However, about 2 months later, Gaia14aae underwent a second outburst, which was detected by {\it Gaia} on 2014 August 11 at {\it G}~=~16.04, during the science commissioning phase. As this was significantly (1.52 mag) brighter than the historic {\it Gaia} magnitude of the source at this position, it was identified and announced as a {\it Gaia} science alert\footnote{http://gaia.ac.uk/selected-gaia-science-alerts}.

Gaia14aae was discovered as part of the {\it Gaia} Science Alerts (GSA) project (Hodgkin et al. in prep.; \citealp{Wyr:12}), which aims to identify such photometric transients in the {\it Gaia} satellite data, and publicly announce their discovery on a rapid timescale. {\it Gaia} is scanning the entire sky at sub-milliarcsecond resolution with precise photometry and astrometry down to a limiting magnitude of {\it G}$\sim$20 \citep[{\it G} is the {\it Gaia} white light bandpass;][]{Jor:10}. Over the five year mission each position on the sky will be observed on average 70 times. These repeated observations of the entire sky mean that alongside the primary science mission of Gaia, to provide spatial, kinematic and physical  parameters for a billion stars in the Milky Way, the satellite will also observe many transient and time-domain phenomena, which will be explored systematically by the GSA project.

Many of the known AM CVn systems display outbursts and super-outbursts in their long-term light curves \citep{Lev:15,Ram:12}, during which they brighten by 3-4 mag over timescales of 1-2 days and last weeks to months. Currently, it appears that about 60 per cent (27/44) of the known AM CVn systems display outbursts \citep{Lev:15}. {\it Gaia} will play an important role in the discovery of new cataclysmic variables (CVs), both in outburst, and also through their decrease in magnitude during eclipses. From pre-launch simulations, we expect that $\sim$1000 new CVs, including a number of evolved systems and AM CVn systems will be found by {\it Gaia} over its mission lifetime.

\section{Observational Data}
\label{sect:data}

\begin{figure*}
\includegraphics[scale=0.65,angle=-90]{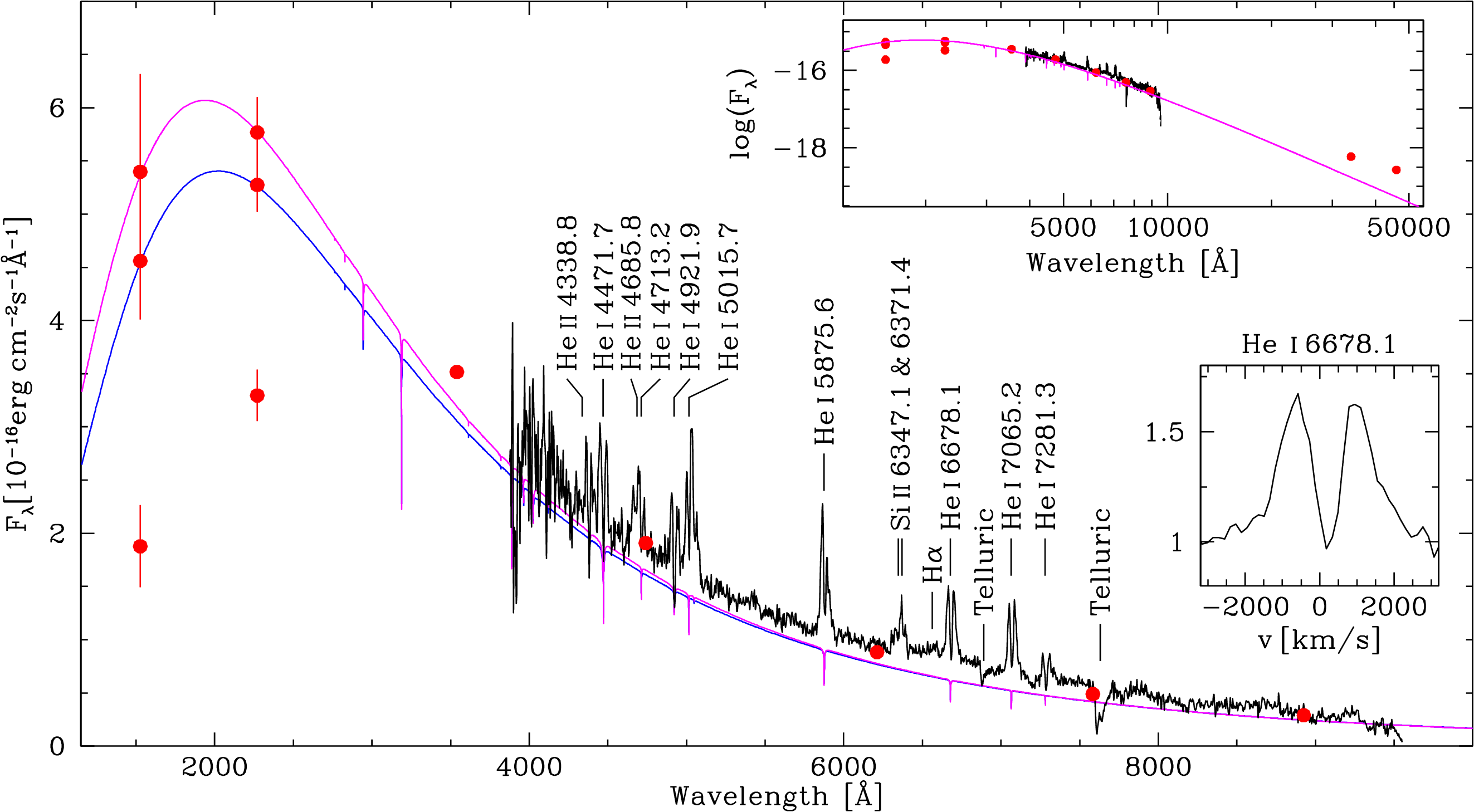}
\caption{\label{WHT} WHT+ACAM spectrum of Gaia14aae taken on 2014 October 13 during quiescence, showing double-peaked He emission and an absence of H lines. The historical \textit{GALEX} and SDSS photometry are also shown as red points; the fainter \textit{GALEX} magnitudes probably cover an eclipse. The blue and magenta lines are $T_\mathrm{eff}$~=~12700\,K and 13100\,K respectively He-atmosphere models fitted to the optical flux and two different epochs of UV flux. The top-right inset shows the spectral energy distribution (SED) fit with the {\it WISE} data included, the bottom-right inset shows a zoom in of the He~{\sc i} $\lambda$6678 line in velocity space.}
\end{figure*}

A 300s long-slit spectrum of Gaia14aae was taken on the night of 2014 October 13 (MJD 56943.88751 at the mid-point of the exposure), when the system had returned to its quiescent state ($i$ = 18.74 $\pm$ 0.02 mag). This spectrum was obtained using the Auxiliary-port camera (ACAM), with the V400 grating, on the 4.2-m William Herschel Telescope (WHT). The data were reduced within {\sc iraf} in the standard fashion. The extracted and calibrated spectrum has a resolution of $\sim$12 \AA\ and a S/N of $\sim$20 in the continuum. The spectrum, plotted in Fig.~\ref{WHT}, shows clear He emission lines, but no detectable H lines. The emission lines are broad (Full width at half maximum (FWHM) = 2415 $\pm$ 100 km/s) and display double-peaked profiles, which are typical of AM CVn stars, revealing the presence of an He-dominated accretion disc.  Based on this spectrum, we classified Gaia14aae as an AM CVn system. The spectral classification and similarities to other eclipses motivated the acquisition of further photometry. The peak velocities of the two emission components are at 800 $\pm$ 50 km/s relative to rest frame, averaged over all detected emission lines. Measurements of the peak separation of the individual double peaked lines all agree within 3$\sigma$ of the average value. We do not see a sharp central spike between the lines, which is observed in many AM CVn stars and thought to originate on the surface of the WD \citep{Mar:99,Mor:03,Roe:07,Roe:09}. This might be due to the low resolution of the spectrum, although it could also be because of the high inclination of this system, as appears to be the case with SDSSJ0926+3624 \citep{Cop:11}.
On the ephemeris given in Section~\ref{sect:results}, the WHT spectrum was taken away from eclipse at phase 0.22. Thus, the absence of the emission spike is not due to the white dwarf eclipse.

The historic optical and infrared fluxes of Gaia14aae in presumed quiescence are also shown in Fig.~\ref{WHT}. The optical fluxes are from the Sloan Digital Sky Survey (SDSS) DR 10 \citep{Aih:11}, while the infrared fluxes are from forced photometry at the SDSS source location \citep{Lan:14} on {\it Wide-Field Infrared Survey Explorer} \citep[{\it WISE};][]{Wri10} images. Ultraviolet (UV) images are also available from {\it Galaxy Evolution Explorer} \citep[\textit{GALEX};][]{Mar:05} DR 6. The \textit{GALEX} archive contains three pairs of FUV \& NUV observations for Gaia14aae, one obtained on 2005 March 9, and the other two on 2007 May 24. All three observations had short exposure times, 143--195\,s. One of the May 2007 observations shows the system at a significantly fainter level than the other two. While our current ephemeris is not sufficiently accurate to establish the orbital phases of the \textit{GALEX} observations, it is most likely that the system was caught close to the eclipse of the primary. The eclipse duration, discussed in Section 3, is 111\,s, shorter than but comparable to the \textit{GALEX}  observations. All fluxes have been corrected for extinction towards the source, $E(B-V)$ = 0.018 \citep{Sch:98}. The absolute flux calibration of the WHT+ACAM spectrum has been scaled to match the SDSS $r$ and $i$ band magnitudes for Gaia14aae.

\begin{table}
\caption{Log of photometric observations of Gaia14aae used in this work.\label{sample_numbers}}
\begin{center}

\begin{tabular}{lccc}
\hline
Observatory 	& Obs. date (UT)		& Filter 	& Exposures (s)				\\
 \hline
{\it Gaia} &  2014 08 11		& $G$&45			\\
ASAS-SN &  2012 - 2015		& $V$&	129$\times$180	\\
Loiano 1.5m Cassini  	& 2014 10 24 		& {\it g} 	& 3$\times$300, 91$\times$30 			\\
Telescope + BFOSC		& 2014 10 25		& {\it g} 	& 135$\times$30 				\\
Bialkow 0.6m, Poland 	& 2014 10 18 		& {\it BV}  	&30$\times$120 				\\
 					& 2014 10 19 		& {\it BV}  	&37$\times$120 				\\
CIECEM 0.35m, Spain 	& 2014 10 21  to		& clear	& 40$\times$180, 8$\times$150 				\\
					& 2014 11 18		& 		& 111$\times$120, 399$\times$90 				\\
\textit{pt5m}, La Palma		& 2014 10 25 		& {\it V}  	& 61$\times$60 				\\	
					& 2014 10 22 		& {\it V}  	& 36$\times$60, 21$\times$120					\\	
0.6m ASV, Serbia		& 2014 10 21 		& $BVRI$  &  6$\times$300 				\\	
Belogradchik AO 0.6m,	& 2014 10 21 		& $BVR$  	& 2$\times$ 300 				\\	
 Bulgaria &  		&   	& 			\\	

Asiago 1.82m Copernico		& 2014 12 11 		& {\it r} 	&169$\times$20 				\\
 					& 2014 12 12 		& {\it g}	&169$\times$20 				\\
4.2m WHT+ACAM		& 2014 12 18 		& $V$  	&491$\times$5 					\\
Mercator & 2015 01 15 & {\it g r+i} & 232$\times$30\\
Catalina  {\it (historic)}	&  2005 - 2014 		& clear 	&107$\times$30 				\\
Pan-STARRS1  {\it (historic)}	 	& 2010 - 2014 		& {\it grizy}&66$\times$30  				\\

 \hline
\end{tabular}
\end{center}
\end{table}

\begin{figure*}
\centering
\includegraphics[scale=1.0,angle=0]{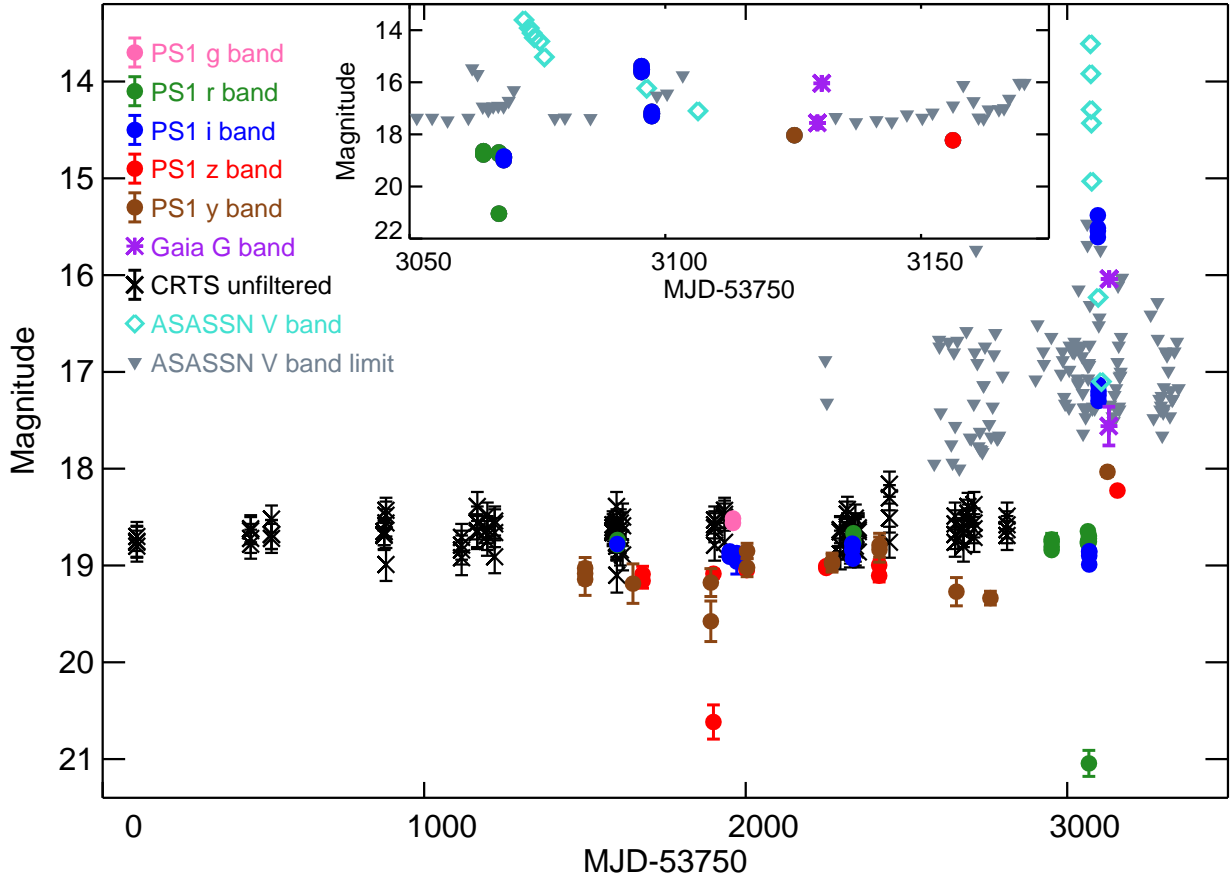}
\caption{\label{CRTS} Historic light curve from CRTS (black crosses) and Pan-STARRS1 (filled circles) spanning eight years of observations. Pan-STARRS1 clearly detected outbursts from Gaia14aae in 2014, and appears to have seen two eclipses. The ASAS-SN and {\it Gaia} detections are shown by the turquoise diamonds (and limits as grey triangles) and a purple stars respectively. This historic light curve begins on 2006 January 5 and ends on 2015 March15.}
\end{figure*}

The initial determination that Gaia14aae was eclipsing was made by the ``Centre for Backyard Astrophysics'' project \citep{Ski:93}; who established a preliminary period for Gaia14aae of 49.7 min \citep{Mig:14}. Following this, an intensive photometric monitoring campaign was undertaken for Gaia14aae at a number of telescopes, as detailed in Table~\ref{sample_numbers}. In addition to this, we searched the databases of the Catalina Real-time Transient Survey \citep[CRTS;][]{Dra:09}, Pan-STARRS-1 \citep[PS1;][]{Mag:13,Sch:12,Ton:12} and ASAS-SN \citep{Sha:14} for pre-discovery images covering the position of Gaia14aae. The cadence of the CRTS data is relatively low, but during those observations no outbursts were observed. The average quiescent magnitude in CRTS for Gaia14aae is 18.64 $\pm$ 0.14 mag. PS1 detected an outburst of Gaia14aae on 2014 July 7, when it reached 15.38 mag in {\it i}-band, compared to 18.74 $\pm$ 0.02 mag in quiescence. Two eclipses of Gaia14aae are also visible in the PS1 data. ASAS-SN has many upper limits for the light curve and detected the decline of the outburst they discovered, as well as some data on the second outburst, but only place limits on the third outburst. The combined light curve for Gaia14aae spanning 8 years of PS1, CRTS, ASAS-SN and {\it Gaia} data is shown in Fig.~\ref{CRTS}. The first {\it Gaia} data point shown is the average of the 1.5 days of data {\it Gaia} had observed before the outburst was discovered. This may already included some of the rise of the outburst and thus be higher than the true historic magnitude.

From the combined light curve, it appears that Gaia14aae underwent at least three outbursts between 2014 June and September. The first outburst was seen by ASAS-SN on 2014 June 14. The second outburst was seen by PS1 in {\it i}-band on 2014 July 7 and ASAS-SN on 2014 July 8. The limits measured by ASAS-SN between 2014 June 20 (6 days after the first outburst) and June 27 (9 days before the second outburst) rule out the possibility that the first and second outburst are in fact one continuing event. The third outburst of the system was caught by {\it Gaia} on 2014 August 13, and is constrained by the {\it Gaia} historic data 1.5 days prior, the PS1 detections of the system in quiescence in {\it i} band 24 days prior and in {\it z} band 7 days after, as well as ASAS-SN limits 1 day after, suggesting this outburst had a short duration.

A number of follow-up studies were conducted. Imaging obtained with the \textit{pt5m}, La Palma (Hardy et al. in prep.) was reduced using the ULTRACAM pipeline \citep{Dhi:07}, while for all other instruments with the exception of WHT+ACAM, the data were debiased and flatfielded using standard techniques. The ACAM data was taken using a small CCD window and a fast readout mode; as no suitable flatfield or bias frames were available, these calibrations have not been applied. However, as we are performing differential photometry over a small area on a single night, this should not affect our results significantly. {\sc  Astrometry.net} \citep{Lan:10} was run on each image, excluding the Cassini+BFOSC, Asiago 1.82m Copernico and WHT+ACAM data, to register it to a common World Coordinate System. {\sc sextractor} \citep{Ber:96} was used to detect, deblend and measure the instrumental magnitudes of all sources in the field. Finally, the list of sources detected in each image was uploaded to the Cambridge Photometry Calibration Server \citep[CPCS\footnote{gsaweb.ast.cam.ac.uk/followup/};][]{Wyr:13}, which calibrates all the data from different telescopes to a common photometric system. To measure the magnitude of Gaia14aae on the BFOSC, Asiago and ACAM images, we used co-located list-driven differential photometry as described in \cite{Irw:07}, using the {\sc CASUTools} package, yielding a precision of 15-18 millimag for Gaia14aae while out of eclipse. The comparison stars were checked and found to be photometrically stable. To correct for light travel times, we converted the MJD (UTC) times of all data to the barycentric dynamical timescale (TDB).

\section{Analysis}
\label{sect:results}

In order to estimate the WD temperature, we assume that the contribution of the accretion disc to the \textit{GALEX} FUV and NUV fluxes is negligible, and fit the three ultraviolet observations with helium-atmosphere models from \citet{Koe:10}, as shown in Fig.~\ref{WHT}. We estimate the contribution of the accretion components from the r-band light curve (discusses below), which is consistent with the assumptions used in the DB model. The two sets of ``bright'' \textit{GALEX} FUV and NUV fluxes are consistent with effective temperature estimates of $T_\mathrm{eff}$~=~12900~$\pm$200~\,K (from $T_\mathrm{eff}$~=~12700\,K, magenta line and 13100\,K, blue line, respectively). DB white dwarfs have very weak lines at such low temperatures, and thus are not detectable given the much stronger emission lines at this resolution, which might explain the lack of broad WD absorption features in the spectrum. Adopting a primary mass of $M_{1}$~=~0.78\,\msun, corresponding to the lower limit from the light curve fit (see below), implies a radius of $R_{1}$~=~7.44$\times10^8$\,cm \citep[using the cooling models of][]{Hol:06}, and hence a distance of 225~$\pm$~10~\,pc. There appears to be an IR excess in the {\it WISE} photometry when compared to the He-atmosphere model. The IR excess is unlikely to be due to outbursts as the {\it WISE} photometry is from observations taken over 2 weeks separated by 6 months. The first set of {\it WISE} data was taken over a period from 2010 July 17--23, while the second set was taken over 2010 December 23--29; there are 7 CRTS measurements during the first set of {\it WISE} observations which constrain the system to be in quiescence. The cause of the {\it WISE} flux excess is unclear.

The ephemeris of Gaia14aae was first determined by fitting a light curve model \citep{Cop:10} to all the photometric data divided into 16 night-long chunks. The model is composed of a WD, accretion disc and a bright-spot where the gas stream hits the disc. The model took into account the finite exposure lengths of the images, including their readout time, by computing over sub-steps in each exposure. We found the ephemeris of Gaia14aae to be\\

\noindent BMJD~(TDB)~=~56980.0557197~(13)~+~0.034519487~(16)~E,\\

\noindent where the zero phase corresponds to the mid-point of the eclipse, based on the time series data from Loiano, Asiago, and WHT.  The time of zero phase was chosen to give minimal correlation between the two fitted parameters and the quoted uncertainties are the 1$\sigma$ errors. At present the estimate of the ephemeris suffers from a few caveats. First, the long (30s or 20s) exposures used for the Loiano and Asiago data, and the small number of eclipse times used (3, 1 and 2 from Loiano, Asiago and WHT, respectively) are not ideal. Second, none of these instruments are built for precise timing and may suffer from systematics.

\begin{figure*}
\includegraphics[width=0.65\linewidth,angle=-90]{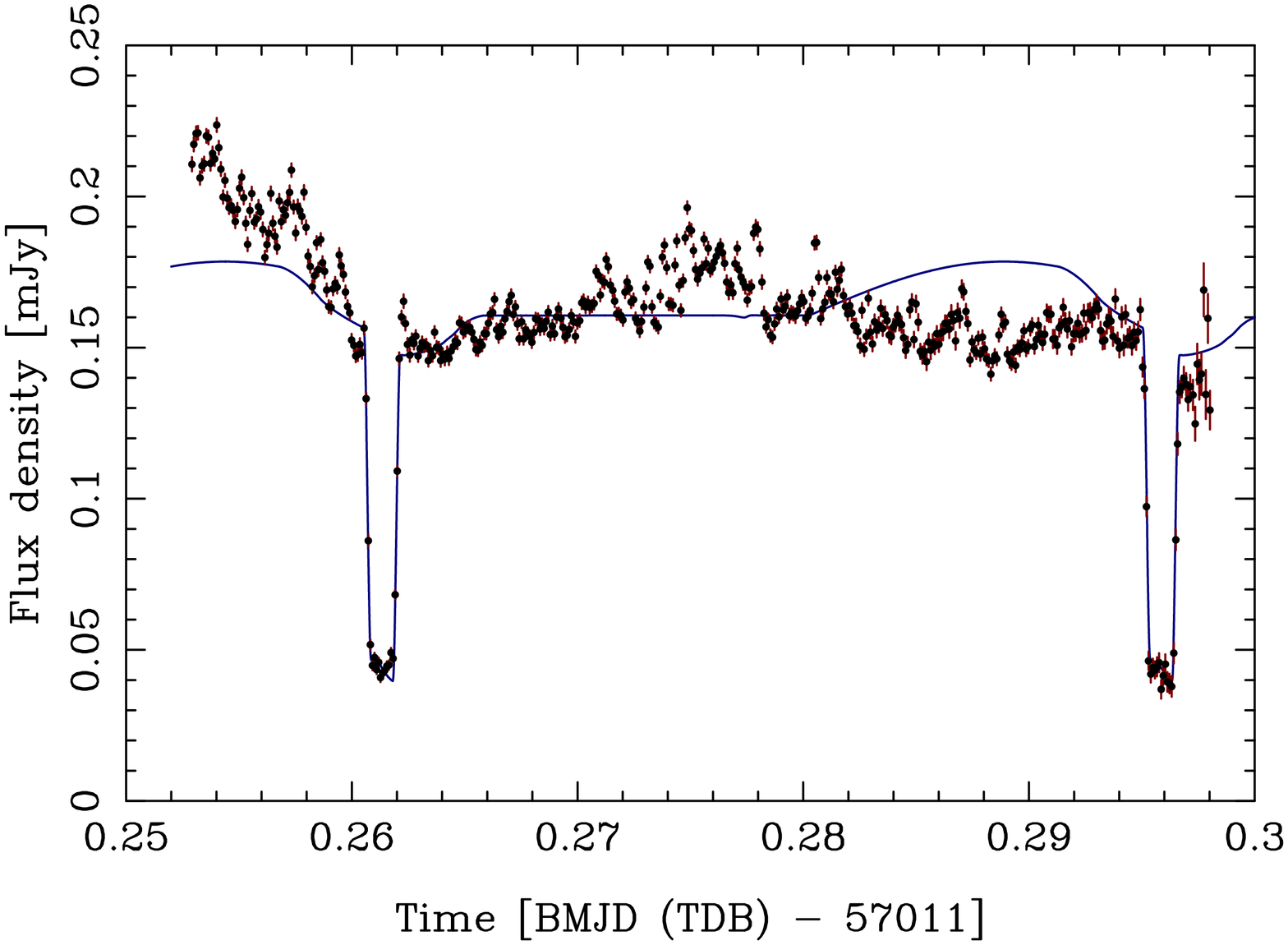}
\includegraphics[width=0.40\linewidth,angle=-90]{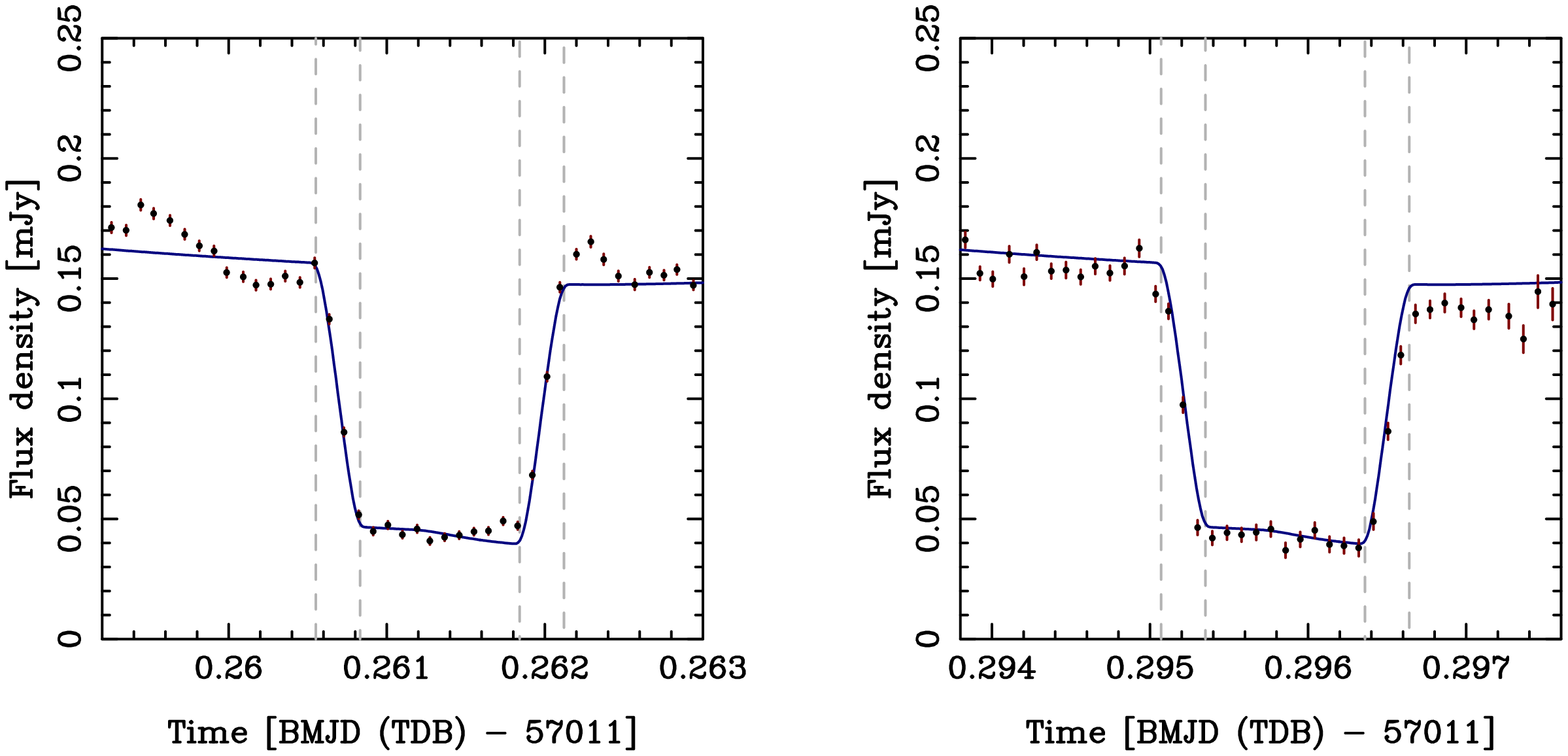}
\caption{\label{WHT_ACAM} Top: Observed {\it r}-band WHT+ACAM light curve for Gaia14aae (points) with the best fitting model (lines) comprising of a WD (which is the main contributor to the light), accretion disc and bright-spot where the gas stream hits the disc. Bottom: Zoom in around the eclipses of the light curve shown above.}
\end{figure*}

We also modelled the high cadence ACAM light curve shown in Fig.~\ref{WHT_ACAM}. The light from this system is dominated by the WD, with a small contribution from the disc and bright spot. It is estimated that the accretion components (bright spot plus accretion disk) contribute $\sim$30 per cent of the $r$-band flux, although of course this component is variable. The pre-eclipse ``hump" which originates in the bright spot seems unusually variable, and sometimes can barely be seen, although this may be due to severe flickering. These are aperiodic brightness variations with characteristic timescales of seconds to minutes \citep{Mid:82}. The amplitude of the flickering exceeds the noise and limits the current model fit. Further observations are required to average the flickering out. The eclipses are sharp-sided and deep, and the mid eclipse depths reach around 2 mag. In order to constrain the scaled white dwarf radius, $r_1$~=~$R_1/a$ (where $R_1$ is radius of the primary and $a$ is the binary separation), we determined the phase of the WD eclipse to be $\Delta \Phi$~=~0.0373~$\pm$~0.0005 from our model fit. The ingress and egress phases were deduced from the parameterised model of the binary fitted to the WHT+ACAM light curve.  This gives us $r_1$ as a function of the mass ratio $q$ and the inclination $i$.  If we then assume a WD mass -- radius relation, we can solve for $M_1$ and $M_2$ using $q$, $r_1$ and the orbital period using Kepler's laws. Here, we assume the relation of P. Eggleton as quoted in \cite{Ver:88}, scaling the relation by a factor of 1.05 to account for the finite temperature of the WD.

There are a range of parameters which fit the current data with our model. The model fits shown in Fig.~\ref{WHT_ACAM} are for $i$~=~88$^{\circ}$. The lower limits to both $M_1$ and $M_2$ correspond to $i$~=~90$\deg$, $q$~=~0.019, $r_1$~=~0.026 and $a$~=~0.413\rsun.  The lower limit on $M_1 \simeq$ 0.782\,\msun\ is consistent with the average mass of WDs in cataclysmic variables \citep{Zor:11}. For $i$~=~90$\deg$, the companion star $M_2$ has a mass of 0.015\,\msun, which is consistent with expectations for a near-zero entropy donor at a period of $\sim$50~min \citep{Del:07}. For a lower inclination model with $i$~=~80$\deg$, $q$~=~0.133, $r_1$~=~0.013, $a$~=~0.488~\rsun, $M_1$ and $M_2$, increase to 1.159 \msun\ and 0.154 \msun\ respectively. From our current data, we are unable to derive a secure value of the mass ratio $q$, due to the flickering and the weak bright-spot. Hence, we cannot select between the low mass, highly degenerate donor stars characteristic of the double WD route as found for $i\sim90\deg$, and more massive hot donors that one might expect from the post-CV route ($i\sim80\deg$). Future high cadence, high S/N observations over multiple orbits might allow us to measure the bright-spot features and break the degeneracy in our derived parameters.

\section{Discussion}

The orbital periods of AM CVn stars are thought to increase as mass is transferred from donor to accretor, leading to a decrease in the rate of mass transfer as the system evolves \citep{Tsu:97,Nel:01}. Thermal instabilities are expected and often observed in AM CVn He accretion discs with intermediate mass-transfer rates, and these are sometimes seen as dwarf nova (DN) type outbursts \citep{Tsu:97}. Intermediate mass-transfer rates are thought to occur for systems with orbital periods of 20~min to $\sim$40~min \citep{Ram:12,Lev:15,Nel:05}. Longer period objects ($P_{\rm orb}~ \gtrsim$~40 min) are thought to have low mass transfer rates and stable cool discs, so that these should not have outbursts, which is mostly confirmed by observations \citep{Ram:12}. However, the low mass transfer rate could also mean that the intervals between outbursts are very long, so we have simply not observed that many outbursts \citep{Lev:15, Kot:12,Can:15}. 

Interestingly, Gaia14aae has experienced three outbursts within only three months, while no outbursts were detected in $\sim$8 years, although we cannot rule out that some could have occurred during gaps in data coverage. Thus to see three outbursts in just a few months, suggests that they are likely to be ``rebrightening" outbursts (also known as echo outbursts), rather than independent events. Multiple rebrightenings are frequently observed in outbursting AM CVn stars and evolved cataclysmic variables \citep[e.g. ][]{Pat:98,She:12,Kat:14,Mey:15}. Echo outbursts are very similar to ``normal" dwarf nova outbursts, except that they happen in quick succession in a system with otherwise few observed outbursts, and they always happen on the decline from a superoutburst.  From Fig.~\ref{CRTS} it can be seen that each outburst reaches a lower peak magnitude than the previous outburst, consistent with echo outbursts, where overall, the target is fading, but it has a few echo outbursts following the superoutburst. In between the rebrightenings it fades to near-quiescence. WZ Sge stars and the outbursting AM CVn stars, such as Gaia14aae, both have low mass transfer rates and extreme mass ratios, which are likely to impact on the duration and frequency of outbursts. \citet{Lev:15} investigates the correlation between orbital period and outburst recurrence time, by extrapolating to rare, long outbursts for long period systems. For our system, with a period of 49.7 mins, they predict outbursts to recur every $\sim$10 years, although this does not consider rebrightenings.

It is somewhat surprising that Gaia14aae shows outbursts at all, because a system with such a long orbital period is expected to have a stable, cool disc \citep{Sol:10}. However, recent studies by \citet{Can:15} and \citet{Kot:12} used a disc instability model and the observed outburst properties of systems, as compiled by \citet{Lev:15}, to find that systems with higher mass accretors, have lower outburst thresholds and are more likely to undergo outbursts. Along with other long period AM CVn stars which experience outbursts, SDSS J090221.35+381941.9 \citep{Kat:14} and CSSJ045019.7+093113 \citep{Wou:13}, these authors suggest that the transition to a stable disc may happen at longer orbital periods in some cases (or perhaps not at all).

 The temperature of the WD implies an accretion rate, if accretion-heated,  of $7-8\times10^{-11}$~\msun\ yr$^{-1}$  for 0.75\msun\ \citep{Tow:09}. Combined with the masses we derive, this accretion rate is more consistent with a degenerate donor \citep{Del:07}, suggesting that the system may have descended from a merging double WD, and that it may have had a much shorter orbital period in the past ($<$10~min).

For comparison, we can compare the accretion rate implied from the WD temperature to the stability criteria of the disc instability model in \citet{Kot:12}. A disc will be stable in the high state if it is too hot and it will be stable in the low state if it is too cold. For the system to be unstable, the accretion rate in the disk must be between the limits for the critical accretion rate for hot ($\dot M^+_{cr}$) and cold ($\dot M^-_{cr}$) stable equilibrium accretion rates. For an inclination of 90$\degree$, with no hydrogen, 98 per cent helium and 2 per cent metals, the upper critical rate is $\dot M^+_{cr}$~=~5.2$\times$10$^{-9}$~\msun~yr$^{-1}$ and lower critical rate is $\dot M^-_{cr}$~=~4.3$\times$10$^{-12}$~\msun~yr$^{-1}$. The WD temperature inferred accretion rate of $7-8\times10^{-11}$~\msun\ yr$^{-1}$  is between the these limits. In fact, for any plausible parameters of the disk instability model, the inferred accretion rate is orders of magnitude below the hot, stable state, and a factor $\sim$20 above the cool, stable state. Thus, Gaia14aae is consistent with the disc instability models for AM CVn stars, since the accretion rate inferred from the WD temperature lies in the unstable regime at this orbital period and mass. Our estimate of the accretion rate could be too high because the WD temperature at the long period of Gaia14aae may be set by simple WD cooling \citep{Bil:06}. However, it would need to be a factor of 20 lower than we estimate to have an accretion rate below the lower critical rate $\dot M^-_{cr}$.

In the future, there is a variety of data which will be essential for fully characterising Gaia14aae. Firstly, more precise, high cadence photometry can be used to average out flickering, which is limiting the analysis of the light curve at present. It is vital to observe the bright spot in the system, as this will precisely pin down the orientation, thus allowing the system parameters to be accurately calculated. Further spectra may allow us to detect the same narrow spikes between the double peaked emission lines that are seen in other AM CVn stars. Combined with the phase from eclipses, this could allow a definitive proof that the spike originates on the accreting WD. Spectra will also provide information on the elements present in the system, useful for understanding the evolutionary history of Gaia14aae. {\it Gaia} will also provide parallax and proper motion, and having an accurate distance to the system will allow system parameters, such as the WD temperature, to be better constrained. Finally, long term precision timing will be needed to detect the expected period change due to  gravitational radiation-driven mass transfer. This effect should cause a progressive delay in the arrival time of eclipses, but it may be at least a decade before this can be detected.

\section{Conclusions}
Gaia14aae was found as a transient in {\it Gaia} data on 2014 August 11. We undertook spectroscopic and photometric follow-up and identify it as an AM CVn system. Gaia14aae is a deeply eclipsing system, with the accreting white dwarf being totally eclipsed on a period of 0.034519~days (49.71~min). It is the third eclipsing AM CVn known, the second in which the white dwarf is eclipsed, and the first in which the white dwarf is totally eclipsed. We detected three outbursts over $\sim$4 months. The orbital period places Gaia14aae at the long period extremum of the outbursting region of the AM CVn distribution. A helium WD model was used to estimate an effective temperature of $\sim$12900~$\pm$~200~K for the white dwarf. We used the contact phases of the WD eclipse to place lower limits of 0.78\msun\ and 0.015\msun\ on the masses of the accretor and donor respectively, which correspond to an inclination of 90$\deg$, a mass ratio of 0.019 and an orbital separation of 0.41\rsun. The deep eclipses shown by Gaia14aae, suggest that future observation have the potential to lead to the most precise parameter determinations of any AM CVn star discovered to date.

\section{Acknowledgements}

We thank the referee. We acknowledge ESA {\it Gaia} (cosmos.esa.int/gaia), DPAC (http://www.cosmos.esa.int/web/gaia/dpac), and the DPAC Photometric Science Alerts Team (gaia.ac.uk/selected-gaia-science-alerts). This work was partly supported by the European Union FP7 programme through ERC grant nos. 320360 and 320964 (WDTracer). TRM, EB and DS acknowledge support from the UK STFC in the form of a Consolidated Grant \#ST/L00073. AH acknowledges support from the Leverhulme Trust through grant RPG-2012-541. MT, LT and PO are partially supported by the PRIN-INAF 2014 Transient Universe: unveiling new types of stellar explosions with PESSTO. The research leading to these results has received funding from the European Union Seventh Framework Programme [FP7/2007-2013] under grant agreement num. 264895. We acknowledge support from the Polish NCN grant 2012/06/M/ST9/00172 to LW, OPTICON FP7 EC grant no 312430 and Polish MNiSW W32/7.PR/2014 grant to LW. TW-SH is supported by the DOE Computational Science Graduate Fellowship, grant number DE-FG02-97ER25308. Support for JLP is provided in part by FONDECYT through the grant 1151445 and by the Ministry of Economy, Development, and Tourism's Millennium Science Initiative through grant IC120009, awarded to The Millennium Institute of Astrophysics, MAS. GD and OV gratefully acknowledge the observing grant support from the Institute of Astronomy and Rozhen NAO BAS; this work is in line with the Projects No 176011, No 176004 and No 176021 supported by the Ministry of Education, Science and Technological Development of the Republic of Serbia. B.S. is supported by NASA through Hubble Fellowship grant HF-51348.001 awarded by the Space Telescope Science Institute, which is operated by the Association of Universities for Research in Astronomy, Inc., for NASA, under contract NAS 5-26555. PAW thanks the National Research Foundation for a research grant. We acknowledge support from the Polish NCN grant 2011/03/B/ST9/02667 to ZK.

We thank the Comit\'e Cient\'ifico Internacional (CCI) of the European Northern Observatory (ENO) in the Canary Islands, for awarding time for this project under the 5\% International Time Programme (ITP).

Pan-STARRS1 is run by the Institute for Astronomy, University of Hawaii, Pan-STARRS Project Office,  Max Planck Institute for Astronomy, Heidelberg, Max Planck Institute for Extraterrestrial Physics, Garching, Johns Hopkins University, Durham University, University of Edinburgh, Queen's University Belfast, Harvard-Smithsonian Center for Astrophysics, Las Cumbres Observatory Global Telescope Network Incorporated, National Central University of Taiwan, Space Telescope Science Institute, NSF Grant AST-1238877, the University of Maryland, and Eotvos Lorand University, Los Alamos National Laboratory,  and NASA Grants NNX08AR22G issued by the Planetary Science Division of the NASA Science Mission Directorate, and grants NNX12AR65G, NNX14AM74G issued through the NEO Observation Program.

The CSS survey is funded by NASA under Grant No. NNG05GF22G issued through the Science Mission Directorate Near-Earth Objects Observations Program. The CRTS survey is supported by the U.S.~National Science Foundation under grants AST-0909182 and AST-1313422. The WHT is operated by the Isaac Newton Group in the Spanish Observatorio del Roque de los Muchachos (Proposal P29). The Copernico 1.82m telescope is operated by INAF Osservatorio Astronomico di Padova. The Cassini 1.52m telescope is operated by INAF Osservatorio Astronomico di Bologna.

\bsp

\label{lastpage}

\end{document}